\documentstyle[12pt]{article}

\begin{document}
\baselineskip=18pt plus 0.2pt minus 0.1pt
\parskip = 6pt plus 2pt minus 1pt
\newcommand{\reseteqnum}{\setcounter{equation}{0}}
\newcommand{\sD}{D\hspace{-0.63em}/}
\newcommand{\al}{\alpha}
\newcommand{\be}{\beta}
\newcommand{\ve}{\varepsilon}
\newcommand{\ep}{\epsilon}
\newcommand{\la}{\lambda}
\newcommand{\La}{\Lambda}
\newcommand{\ga}{\gamma}
\newcommand{\si}{\sigma}
\newcommand{\de}{\delta}
\newcommand{\bx}{\mbox{\boldmath $x$}}
\newcommand{\bX}{\mbox{\boldmath $X$}}
\newcommand{\bY}{\mbox{\boldmath $Y$}}
\newcommand{\D}{\partial}
\newcommand{\rhovev}{(\rho v)^2}
\newcommand{\mr}{\frac{g^2}{\lambda}}
\begin{titlepage}
\title{
\hfill
\parbox{4cm}{\normalsize KUCP-0078\\KUNS-1346\\HE(TH)~95/08\\hep-th/9507111}\\
\vspace{1cm}
Valley Instanton in the Gauge-Higgs System
}
\author{Hideaki Aoyama\thanks{e-mail address:
\tt aoyama@phys.h.kyoto-u.ac.jp}
\\
{\normalsize\sl Department of Fundamental Sciences, FIHS,
Kyoto University, Kyoto 606-01, Japan}
\vspace{0.5cm}
\\
Toshiyuki Harano\thanks{e-mail address: \tt harano@gauge.scphys.kyoto-u.ac.jp}
\\
{\normalsize\sl Department of Physics, Kyoto University, Kyoto 606-01, Japan}
\vspace{0.5cm}
\\
Masatoshi Sato\thanks{e-mail address: \tt msato@gauge.scphys.kyoto-u.ac.jp}
\\
{\normalsize\sl Department of Physics, Kyoto University, Kyoto 606-01, Japan}
\vspace{0.5cm}
\\
Shinya Wada\thanks{e-mail address: \tt shinya@phys.h.kyoto-u.ac.jp}
\\
{\normalsize\sl Graduate School of Human and Environmental Studies,
Kyoto University}\\
{\normalsize\sl Kyoto 606-01, Japan}}
\date{\normalsize July, 1995}
\maketitle
\thispagestyle{empty}

\begin{abstract}
\normalsize
The instanton configuration in the SU(2)-gauge system with a Higgs doublet
is constructed by using the new valley method.
This method defines the configuration by an extension of the field equation and
allows the exact conversion of the quasi-zero eigenmode
to a collective coordinate.
It does not require ad-hoc constraints used
in the current constrained instanton method and provides
a better mathematical formalism than the constrained instanton method.
The resulting instanton, which we call ``valley instanton'',
is shown to have desirable behaviors.
The result of the numerical investigation is also presented.
\end{abstract}
\end{titlepage}

\newpage
\reseteqnum
\section{Introduction}
The instantons in the gauge-Higgs system play important
roles in the investigation of the nonperturbative aspects of the theory.
One of such nonperturbative phenomena is the baryon number violation process
in the electroweak theory \cite{thooft,manton}.
The tunneling between different vacua, which may not be
negligible in the TeV region, results in the
violation of the baryon number and lepton number
conservation via anomalies in the respective currents.
This process is best evaluated in the instanton method \cite{Ring,Esp}
and its variations \cite{AK}.
The other is the investigation of the SUSY theories, where
the instanton effects are essential in the symmetry breaking
\cite{NSVZ,ADS,Yung,AKMRV}.

Construction of instanton in these theories is not
trivial due to the well-known scaling argument,
which shows that among all the configurations of finite Euclidean
action, only the zero-radius configuration can
be a solution of equation of motion.
Consequently, all of the evaluation mentioned above used
the so-called constrained instanton formalism \cite{Aff}.
In this formalism, one introduces a constraint in the system
so that the finite-radius configuration can be a solution
of the equation of motion under the constraint.
By integrating over the constraint parameter, one hopes to
recover the original functional integral.
One problem about this method is that its validity
depends on the choice of the constraint:
Since in practice one only does the Gaussian integration
around the solution under the constraint, the degree of
approximation depends on the way constraint is introduced.
Unfortunately, no known criterion guarantees the
effectiveness of the approximation.

This situation could be remedied once one realizes that
what we have near the point-like (true) instanton is the valley.
The trajectory along the valley bottom should correspond to
the scaling parameter, or the radius parameter of the instanton.
As such, the finite-size instanton can be defined as
configurations along the valley trajectory.
Furthermore, in practical applications we need to incorporate
the contribution of the instanton-anti-instanton valley \cite{Yung,KR}.
Thus treating a single instanton as a configuration on the valley
provides a means of unifying the approximation scheme.
We dub these configurations ``valley instanton''.

One convenient way to define the valley trajectory
is to use the new valley method \cite{AKnv,aw}.
Denoting all the bosonic fields in the theory by  $\phi_\alpha(x)$,
the new valley equation for the
theory with action $S[\phi_\alpha]$ is written as follows;
\begin{eqnarray}
\sum_\beta \int d^4 y
{\delta^2 S \over \delta\phi_{\alpha}(x) \delta\phi_{\beta}(y)}
F_{\beta}(y) = \lambda F_{\alpha}(x),
\hskip 4mm
F_\alpha (x) = {\delta S \over \delta\phi_{\alpha}(x)},
\label{eq:gennv}
\end{eqnarray}
where $\lambda$ is the smallest eigenvalue of
$\delta^2 S / \delta \phi \delta \phi$.
The eigenvalue $\lambda$ also plays the role
of the parameter on the valley trajectory.
The field $F_\alpha (x)$ is an auxiliary field that
can be eliminated, but is convenient for later analysis.

Unlike the streamline method \cite{balyun} used before, the new valley
equation (\ref{eq:gennv})
is a local equation in the functional space, and
does not require any boundary condition for the valley itself.
Also, it is an extension of the equation of motion,
which can be analyzed by using the conventional methods.
The strength of the method lies in the fact that
it defines the valley trajectory such that the
eigenvalue used in the new valley equation is {\sl completely} removed
from the Gaussian integration.
In other words, by choosing the eigenvalue in the
new valley equation one can remove the unwanted negative, zero
or pseudo zero modes from the Gaussian integral.
In this sense, it defines a method of extending the
ordinary collective coordinate method to the case of
negative or psuedo-zero modes.

In the following, we will write down the new valley equation
for the gauge-Higgs system and present the analytical and
numerical evaluation of the solutions to reveal the properties of
the valley instanton.

\section{New-valley equation in the gauge-Higgs system}
We consider the SU(2) gauge theory with one scalar Higgs doublet,
which has the following action $S=S_g+S_h$;
\begin{eqnarray}
&&S_g=\frac{1}{2 g^2} \int d^4 x\ {\rm tr} F_{\mu \nu}  F_{\mu \nu},
\label{gaugeaction}
\\
&&S_h=\frac{1}{ \lambda}\int d^4 x\left\{\left(D_{\mu} H\right)^{\dagger}
\left(D_{\mu} H\right)+
\frac{1}{8}\left(H^{\dagger}H-v^2\right)^2\right\},
\label{higgsaction}
\end{eqnarray}
where $F_{\mu\nu}=\partial_{\mu} A_{\nu}-\partial_{\nu}
A_{\mu}-i\left[A_{\mu},A_{\nu}\right]$ and
$D_{\mu}=\partial_{\mu}-iA_{\mu}$.
The masses of the gauge boson and the Higgs boson are given by,
\begin{eqnarray}
&&m_{_W}=\sqrt{\frac{g^2}{2\lambda}} v,\quad m_{_H}=\frac{1}{\sqrt{2}} v.
\end{eqnarray}
The valley equation for this system is given by,
\begin{eqnarray}
&&\frac{\delta^2 S}{\delta A_{\mu}\delta A_{\nu}}F^A_{\nu}
+\frac{\delta^2 S}{\delta A_{\mu}\delta H^{\dagger}}F^H
+\frac{\delta^2 S}{\delta A_{\mu}\delta H}F^{H\dagger}
=\lambda_e F^A_{\mu}\nonumber,
\\
&&\frac{\delta^2 S}{\delta H^{\dagger}\delta A_{\mu}}
F^A_{\mu}+
\frac{\delta^2 S}{\delta H^{\dagger}\delta H}F^{H\dagger}
+\frac{\delta^2 S}{\delta H^{\dagger}\delta H^{\dagger}}F^H
=\lambda_e F^{H\dagger},
\label{valley}
\\
&&F^A_{\mu}=\frac{\delta S}{\delta A_{\mu}},
\quad F^H =\frac{\delta S}{\delta H},\nonumber
\end{eqnarray}
where the integration over the space-time is implicit.
The valley is parametrized by the eigenvalue $\lambda_e$ which
is identified with the zero mode corresponding to the scale invariance
 in the massless limit, $v\rightarrow 0$.

To simplify the valley equation, we adopt the following ansatz;
\begin{eqnarray}
A_{\mu}(x)=\frac{x_{\nu}\bar{\sigma}_{\mu\nu}}{x^2}\cdot 2a(r),
\quad H(x)=v\left( 1-h(r)\right)\eta,
\label{ansatz}
\end{eqnarray}
where $\eta$ is a constant isospinor, and $a$ and
$h$ are real dimensionless functions of dimensionless variable
$r$, which is defined by $r=\sqrt{x^2}/\rho$.
We have introduced the scaling parameter $\rho$ so that we adjust the
radius of valley instanton as we will see later. The tensor structure
in (\ref{ansatz}) is the same as that of the instanton  in the singular
gauge. Inserting this ansatz to (\ref{valley}), the structure
of $F^A_{\mu}$ and $F^{H\dagger}$ is
determined as the following;
\begin{eqnarray}
F^A_{\mu}(x)=\frac{x_{\nu}\bar{\sigma}_{\mu\nu}}{x^2}\cdot
\frac{2v^2}{\lambda}f^a(r),
&&F^{H\dagger}(x)=-\frac{v^3}{\lambda}f^h(r)\eta.
\label{fansatz}
\end{eqnarray}
By using this ansatz, (\ref{ansatz}) and (\ref{fansatz}), the
valley equation (\ref{valley}) is reduced to the following;
\begin{eqnarray}
&&-\frac{1}{r}\frac{d}{dr}\left(r\frac{da}{dr}\right)
+\frac{4}{r^2}a(a-1)(2a-1)
+\frac{g^2}{2\lambda}\rhovev a(1-h)^2=\mr(\rho
v)^2 f^a,
\label{eqn:va}\\
&&-\frac{1}{r^3}\frac{d}{dr}\left(r^3\frac{dh}{dr}\right)
+\frac{3}{r^2}(h-1)a^2
+\frac{1}{4}\rhovev h(h-1)(h-2)=\rhovev f^h,
\label{eqn:vh}\\
&&-\frac{1}{r}\frac{d}{dr}\left(r\frac{df^a}{dr}\right)
+\frac{4}{r^2}(6a^2-6a+1)f^a+\frac{g^2}{2\lambda}\rhovev (h-1)^2f^a
\nonumber
\\
&&\hspace{30ex}+\mr\rhovev a(h-1)f^h=\mr\rhovev \nu f^a,
\label{eqn:vfa}\\
&&-\frac{1}{r^3}\frac{d}{dr}\left(r^3\frac{df^h}{dr}\right)
+\frac{3a^2}{r^2}f^h+\frac{1}{4}\rhovev (3h^2-6h+2)f^h
\nonumber
\\
&&\hspace{30ex}+\frac{6a}{r^2}(h-1)f^a=\rhovev \nu f^h,
\label{eqn:vfh}
\end{eqnarray}
where $\nu$ is defined as $\lambda_e=v^2\nu/\lambda$.

In the massless limit, $\rho v \rightarrow 0$, (\ref{eqn:va}) and
(\ref{eqn:vh}) reduce to the equation of motion and (\ref{eqn:vfa})
and (\ref{eqn:vfh}) to the equation for the zero-mode fluctuation around the
instanton solution. The solution of this set of equations is the following;
\begin{eqnarray}
\begin{array}{cc}
\displaystyle a_{0}=\frac{1}{1+r^{2}},
&\displaystyle h_{0}=1-\left(\frac{r^{2}}{1+r^{2}}\right)^{1/2},
\\
\displaystyle f^a_{0}=\frac{2Cr^{2}}{(1+r^{2})^{2}},
&\displaystyle f^h_{0}=\frac{Cr}{(1+r^{2})^{3/2}},
\end{array}
\label{eq:a0}
\end{eqnarray}
where $C$ is an arbitrary function of $\rho v$.
Note that $a_{0}$ is an instanton solution in the singular gauge and
$h_{0}$ is a Higgs configuration in the instanton background \cite{thooft}.
We have adjusted the scaling parameter $\rho$ so that the radius of the
instanton solution is unity. The mode solutions
$f^a_{0}$ and $f^h_{0}$ are obtained from $\partial a_0/\partial \rho$ and
$\partial h_0/\partial \rho$, respectively.

\section{Analytic construction of the valley instanton}
In this section we will construct the valley instanton analytically.
When $\rho v=0$, it is given by the ordinary instanton configuration $a_{0}$,
$h_{0}$, $f^{a}_{0}$ and $f^{h}_{0}$.
When $\rho v$ is small but not zero, it is expected that small $\rho v$
corrections appear in the solution.
On the other hand, at large distance from the core of the valley
instanton, this solution is expected to decay exponentially, because
gauge boson and Higgs boson are massive.
Therefore, the solution is similar to the
instanton near the origin and decays exponentially in the asymptotic region.
In the following, we will solve the valley equation in both region
and analyze the connection in the intermediate region.
In this manner we will find the solution.

In the asymptotic region, $a$, $h$, $f^{a}$
and $f^{h}$ become small and the valley equation can be linearlized;
\begin{eqnarray}
&&-\frac{1}{r}\frac{d}{dr}\left(r\frac{da}{dr}\right)
+\frac{4}{r^{2}}a+\frac{g^{2}}{2\lambda}(\rho v)^{2}a
=\frac{g^{2}}{\lambda}(\rho v)^{2}f^{a},
\label{eqn:la}\\
&&-\frac{1}{r^{3}}\frac{d}{dr}\left(r^{3}\frac{dh}{dr}\right)
+\frac{1}{2}(\rho v)^{2}h=(\rho v)^{2}f^{h},
\label{eqn:lh}\\
&&-\frac{1}{r}\frac{d}{dr}\left(r\frac{df^{a}}{dr}\right)
+\frac{4}{r^{2}}f^{a}+\frac{g^{2}}{2\lambda}(\rho v)^{2}f^{a}
=\frac{g^{2}}{\lambda}(\rho v)^{2}\nu f^{h},
\label{eqn:lfa}\\
&&-\frac{1}{r^{3}}\frac{d}{dr}\left(r^{3}\frac{df^{h}}{dr}\right)
+\frac{1}{2}(\rho v)^{2}f^{h}=(\rho v)^{2}\nu f^{h}.
\label{eqn:lfh}
\end{eqnarray}
The solution of this set of equations is
\begin{eqnarray}
&&a(r)=C_{1}\,r\frac{d}{dr}G_{\rho m_{_W}}(r)+\frac{1}{\nu}f^{a}(r),\\
&&h(r)=C_{2}\,G_{\rho m_{_H}}(r)+\frac{1}{\nu}f^{h}(r),\\
&&f^{a}(r)=C_{3}\,r\frac{d}{dr}G_{\rho \mu_{_W}}(r),\\
&&f^{h}(r)=C_{4}\,G_{\rho \mu_{_H}}(r),
\end{eqnarray}
where $C_{i}$ are arbitrary functions of $\rho v$ and $\mu_{_{W,\,H}}$
are defined as
$\mu_{_{W,\,H}}=m_{_{W,\,H}}\sqrt{1-2\nu}$.
The function $G_{\mu}(r)$ is
\begin{equation}
G_{\mu}(r)=\frac{\mu K_{1}(\mu r)}{(2\pi)^{2}r},
\end{equation}
where $K_{1}$ is a modified Bessel function.
As was expected above, these solutions decay exponentially at
infinity and
when $r\ll(\rho
v)^{-1}$ they have the series expansions;
\begin{eqnarray}
&&a(r)=\frac{C_{1}}{(2\pi)^{2}}
\left[-\frac{2}{r^{2}}+\frac{1}{2}(\rho m_{_W})^{2}+\cdots\right]
+\frac{C_{3}}{\nu(2\pi)^{2}}
\left[-\frac{2}{r^{2}}+\frac{1}{2}(\rho \mu_{_W})^{2}+\cdots\right],
\label{eqn:aa}\\
&&h(r)=\frac{C_{2}}{(2\pi)^{2}}
\left[\frac{1}{r^{2}}+\frac{1}{2}(\rho m_{_H})^{2}\ln(\rho m_{_H}rc)
+\cdots\right]\nonumber\\
&&\hspace{30ex}+\frac{C_{4}}{\nu(2\pi)^{2}}
\left[\frac{1}{r^{2}}+\frac{1}{2}(\rho \mu_{_H})^{2}\ln(\rho \mu_{_H}rc)
+\cdots\right],
\label{eqn:ah}\\
&&f^{a}(r)=\frac{C_{3}}{(2\pi)^{2}}
\left[-\frac{2}{r^{2}}+\frac{1}{2}(\rho \mu_{_W})^{2}+\cdots\right],
\label{eqn:afa}\\
&&f^{h}(r)=\frac{C_{4}}{(2\pi)^{2}}
\left[\frac{1}{r^{2}}+\frac{1}{2}(\rho \mu_{_H})^{2}\ln(\rho \mu_{_H}rc)
+\cdots\right],
\label{eqn:afh}
\end{eqnarray}
$c$ being a numerical constant $e^{\gamma-1/2}/2$, where $\gamma$ is
the Euler's constant.

Near the origin, we expect that the valley instanton is similar to the
ordinary instanton.
Then the following replacement of the field variables is convenient;
$a=a_{0}+(\rho v)^{2}\hat{a}$,
$h=h_{0}+(\rho v)^{2}\hat{h}$,
$f^{a}=f^{a}_{0}+(\rho v)^{2}\hat{f^{a}}$,
$f^{h}=f^{h}_{0}+(\rho v)^{2}\hat{f^{h}}$.
If we assume
$a_{0}\gg(\rho v)^{2}\hat{a}$,
$h_{0}\gg(\rho v)^{2}\hat{h}$,
$f^{a}_{0}\gg(\rho v)^{2}\hat{f^{a}}$
and $f^{h}_{0}\gg(\rho v)^{2}\hat{f^{h}}$,
the valley equation becomes
\begin{eqnarray}
&&-\frac{1}{r}\frac{d}{dr}\left(r\frac{d\hat{a}}{dr}\right)
+\frac{4}{r^{2}}(6a_{0}^{2}-6a_{0}+1)\hat{a}
+\frac{g^{2}}{2\lambda}a_{0}(h_{0}-1)^{2}
=\frac{g^{2}}{\lambda}f^{a}_{0},
\label{eqn:ha}\\
&&-\frac{1}{r^{3}}\frac{d}{dr}\left(r^{3}\frac{d\hat{h}}{dr}\right)
+\frac{3}{r^{2}}a_{0}^{2}\hat{h}+\frac{6}{r^{2}}(h_{0}-1)a_{0}\hat{a}
+\frac{1}{4} h_{0}(h_{0}-1)(h_{0}-2)=f^{h}_{0},
\label{eqn:hh}\\
&&-\frac{1}{r}\frac{d}{dr}\left(r\frac{d\hat{f^{a}}}{dr}\right)
+\frac{4}{r^{2}}(6a_{0}^{2}-6a_{0}+1)\hat{f^{a}}
+\frac{24}{r^{2}}(2a_{0}-1)f^{a}_{0}\hat{a}\nonumber\\
&&\hspace{28ex}+\frac{g^{2}}{2\lambda}(h_{0}-1)^{2}f^{a}_{0}
+\frac{g^{2}}{\lambda}a_{0}(h_{0}-1)f^{h}_{0}
=\frac{g^{2}}{\lambda}\nu f^{a}_{0},
\label{eqn:hfa}\\
&&-\frac{1}{r^{3}}\frac{d}{dr}\left(r^{3}\frac{d\hat{f^{h}}}{dr}\right)
+\frac{3}{r^{2}}a_{0}^{2}\hat{f^{h}}
+\frac{6}{r^{2}}a_{0}f^{h}_{0}\hat{a}
+\frac{1}{4}(3h_{0}^{2}-6h_{0}+2)f^{h}_{0}\nonumber
\\
&&\hspace{20ex}+\frac{6}{r^{2}}a_{0}(h_{0}-1)\hat{f^{a}}
+\frac{6}{r^{2}}(h_{0}-1)f^{a}_{0}\hat{a}
+\frac{6}{r^{2}}a_{0}f^{a}_{0}\hat{h}
=\nu f^{h}_{0}.
\label{eqn:hfh}
\end{eqnarray}
To solve this eqation, we introduce solutions of the following
equations;
\begin{equation}
\begin{array}{l}
\displaystyle
-\frac{1}{r}\frac{d}{dr}\left(r\frac{d\varphi_{a}}{dr}\right)
+\frac{4}{r^{2}}(6a_{0}^{2}-6a_{0}+1)\varphi_{a}=0,
\\
\\
\displaystyle
-\frac{1}{r^{3}}\frac{d}{dr}\left(r^{3}\frac{d\varphi_{h}}{dr}\right)
+\frac{3}{r^{2}}a_{0}^{2}\varphi_{h}=0.
\label{eqn:zh}
\end{array}
\end{equation}
They are given as,
\begin{equation}
\varphi_{a}=\frac{r^{2}}{(1+r^{2})^{2}},
\quad\varphi_{h}=\left(\frac{r^{2}}{1+r^{2}}\right)^{1/2}.
\end{equation}
Using these solutions, we will integrate the valley equation.
We multiply (\ref{eqn:ha}) and (\ref{eqn:hfa}) by $r\varphi_{a}$, and
multiply (\ref{eqn:hh}) and (\ref{eqn:hfh}) by $r^{3}\varphi_{h}$ then
integrate them from $0$ to $r$.
Integrating by parts and using (\ref{eqn:zh}), we obtain
\begin{eqnarray}
&&-\varphi_{a}r\frac{d\hat{a}}{dr}+\frac{d\varphi_{a}}{dr}r\hat{a}
=\frac{g^{2}}{\lambda}\int_{0}^{r}dr' r'\varphi_{a}
\left[f^{a}_{0}-\frac{1}{2}a_{0}(h_{0}-1)^{2}\right],
\label{eqn:dha}\\
&&-\varphi_{h}r^{3}\frac{d\hat{h}}{dr}+\hat{h}r^{3}\frac{d\varphi_{h}}{dr}
=\int_{0}^{r}dr' r'^{3}\varphi_{h}
\left[f^{h}_{0}-\frac{1}{4}h_{0}(h_{0}-1)(h_{0}-2)
-\frac{6}{r'^{2}}(h_{0}-1)a_{0}\hat{a}\right],
\label{eqn:dhh}\\
&&-\varphi_{a}r\frac{d\hat{f^{a}}}{dr}
+\frac{d\varphi_{a}}{dr}r\hat{f^{a}}\nonumber\\
&&\hspace{4ex}=\int_{0}^{r}dr' r'\varphi_{a}
\left[\frac{g^{2}}{\lambda}\nu f^{a}_{0}
-\frac{24}{r'^{2}}(2a_{0}-1)f^{a}_{0}\hat{a}
-\frac{g^{2}}{2\lambda}(h_{0}-1)^{2}f^{a}_{0}
-\frac{g^{2}}{\lambda}a_{0}(h_{0}-1)f^{h}_{0}\right],
\label{eqn:dhfa}\\
&&-\varphi_{h}r^{3}\frac{d\hat{f^{h}}}{dr}
+\hat{h}r^{3}\frac{d\varphi_{h}}{dr}\nonumber\\
&&\hspace{4ex}=\int_{0}^{r}dr' r'^{3}\varphi_{h}
\left[\nu f^{h}_{0}-\frac{6}{r'^{2}}a_{0}f^{a}_{0}\hat{a}
-\frac{1}{4}(3h_{0}^{2}-6h_{0}+2)f^{h}_{0}\right.\nonumber\\
&&\hspace{31ex}\left.-\frac{6}{r'^{2}}a_{0}(h_{0}-1)\hat{f^{a}}
-\frac{6}{r'^{2}}(h_{0}-1)f^{a}_{0}\hat{a}
-\frac{6}{r'^{2}}a_{0}f^{a}_{0}\hat{h}\right].
\label{eqn:dhfh}
\end{eqnarray}

First we will find $\hat{a}$.
The right-handed side of (\ref{eqn:dha}) is proposional to $(C-1/4)$
and when $r$ goes to infinity this approaches a constant.
At $r\gg1$, (\ref{eqn:dha}) becomes
\begin{equation}
-\frac{1}{r}\frac{d\hat{a}}{dr}-\frac{2}{r^{2}}\hat{a}
=\frac{1}{3}\frac{g^{2}}{\lambda}\left(C-\frac{1}{4}\right).
\end{equation}
Then at $r\gg1$, $\hat{a}(r)$ is propotional to
$(C-1/4)r^{2}$ and $a(r)$ becomes
\begin{equation}
a=\frac{1}{r^{2}}
-\frac{(\rho v)^{2}g^{2}}{12\lambda}\left(C-\frac{1}{4}\right)r^{2}+\cdots.
\end{equation}
To match this with (\ref{eqn:aa}), it must be hold that $C=1/4$ when
$\rho v=0$.
When $C=1/4$, the right-handed sides of (\ref{eqn:dha}) vanishes and
$\hat{a}$ satisfy $-\varphi_{a}d\hat{a}/dr+\hat{a}d\varphi_{a}/dr=0$.
Hence $\hat{a}$ is $\hat{a}=D\,\varphi_{a}$, where $D$ is a constant.
Identifing $a_{0}+(\rho v)^{2}\hat{a}$ with (\ref{eqn:aa}) again at $r\gg1$,
we find that
$C_{1}+C_{3}/\nu=-2\pi^{2}$ and $C_{3}=-\pi^{2}$ at $\rho v=0$.
In the same manner, $\hat{h}$,
$\hat{f^{a}}$ and $\hat{f^{h}}$ are obtained.
At $r\gg1$, we find
\begin{eqnarray}
&&\hat{h}={\rm const.}+\cdots,
\nonumber\\
&&\hat{f^{a}}=\frac{g^{2}}{48\lambda}\left(\frac{1}{4}-\nu\right)r^{2}
-\frac{g^{2}}{16\lambda}(1-2\nu)+\cdots,
\\
&&\hat{f^{h}}=\frac{1}{16}(1-2\nu)\ln r+\cdots.
\nonumber
\end{eqnarray}
Here ${\rm const.}$ is a constant of integration.
Comparing (\ref{eqn:ah})-(\ref{eqn:afh}) with
them, we find that $C_{2}+C_{4}/\nu=2\pi^{2}$,
$C_{4}=\pi^{2}$ and $\nu=1/4$ at $\rho v=0$.

Now we have obtained the solution of the new valley equation.
Near the origin of the valley instanton, it is given as $a=a_{0}$,
$h=h_{0}$, $f^{a}=f^{a}_{0} $ and $f^{h}=f^{h}_{0}$.
As $r$ becomes large, the correction terms become important;
\begin{equation}
\begin{array}{l}
\displaystyle
a=\frac{1}{r^{2}}+\cdots,
\hspace{15ex}h=\frac{1}{2r^{2}}-\frac{(\rho v)^{2}}{16}\ln2+\cdots,
\\
\\
\displaystyle
f^{a}=\frac{1}{2r^{2}}-\frac{g^{2}(\rho v)^{2}}{32\lambda}+\cdots,
\quad f^{h}=\frac{1}{4r^{2}}-\frac{(\rho v)^{2}}{32}\ln r+\cdots,
\end{array}
\end{equation}
and finally it is given by (\ref{eqn:aa})-(\ref{eqn:afh}), where
$C_{1}=2\pi^{2}$,
$C_{2}=-2\pi^{2}$, $C_{3}=-\pi^{2}$, $C_{4}=\pi^{2}$ and $\nu=1/4$ at
$\rho v=0$.
Let us make a brief comment about the consistecy of our analysis.
Until now, we have implicitly assumed that there exists an overlapping
region where both (\ref{eqn:la})-(\ref{eqn:lfh}) and
(\ref{eqn:ha})-(\ref{eqn:hfh}) are valid.
Using the above solution, it is found that (\ref{eqn:la})-(\ref{eqn:lfh})
are valid when $r\gg (\rho v)^{-1/2}$ and (\ref{eqn:ha})-(\ref{eqn:hfh})
are valid when $r\ll (\rho v)^{-1}$.
If $\rho v$ is small enough,
there exists the overlapping region $(\rho v)^{-1/2}\ll r\ll (\rho
v)^{-1}$.
Then our analysis is consistent.

The action of the valley instanton can be caluculated using the above
solution.
Rewriting the action in terms of  $a$ and $h$, we find
\begin{eqnarray}
&&S_{g}=\frac{12\pi^{2}}{g^{2}}
\int_{0}^{\infty}\frac{dr}{r}\left\{
\left( r\frac{da}{dr}\right)^{2}+4a^{2}(a-1)^{2}
\right\},
\\
&&S_{h}=\frac{2\pi^{2}}{\lambda}(\rho v)^{2}\int_{0}^{\infty}
r^{3}dr\left\{
\left(\frac{dh}{dr}\right)^{3}+\frac{3}{r^{2}}(h-1)^{2}a^{2}
+\frac{1}{8}(\rho v)^{2}h^{2}(h-2)^{2}\right\}.
\end{eqnarray}
Substituting the above solution for $S$, we obtain
\begin{equation}
\label{eqn:aaction}
S=\frac{8\pi^{2}}{g^{2}}
    +\frac{2\pi^{2}}{\lambda}(\rho v)^{2}+O((\rho v)^{4}\ln(\rho v)).
\end{equation}
The leading contribution $8\pi^{2}/g^{2}$ comes from $S_{g}$ for $a_{0}$,
which is the action of the instanton, and the next-to-leading
contribution comes from $S_{h}$ for $a_{0}$ and $h_{0}$.

\section{Numerical analysis}
\input epsf
In this section, we solve the valley equation
(\ref{eqn:va})-(\ref{eqn:vfh})
numerically.
We need a careful discussion for solving the valley equation
(\ref{eqn:va})-(\ref{eqn:vfh}):
Since the solution must be regular at the origin, we assume the following
expansions;
\begin{eqnarray}
\begin{array}{cc}
\displaystyle a(r)=\sum_{n=0}^{\infty} a_{(n)} r^n,
&\displaystyle h(r)=\sum_{n=0}^{\infty} h_{(n)} r^n,
\\
\displaystyle f^a(r)=\sum_{n=0}^{\infty} f^a_{(n)} r^n,
&\displaystyle f^h(r)=\sum_{n=0}^{\infty} f^h_{(n)} r^n,
\end{array}
\label{expand}
\end{eqnarray}
for $r \ll 1$.
Inserting (\ref{expand}) to (\ref{eqn:va})-(\ref{eqn:vfh}), we obtain
\begin{eqnarray}
&&a_{(0)}=1,h_{(0)}=1,f^a_{(0)}=0,f^h_{(0)}=0,
\nonumber
\\
&&a_{(1)}=0,f^a_{(1)}=0,
\\
&&h_{(2)}=0,f^h_{(2)}=0.
\nonumber
\end{eqnarray}
The coefficients $a_{(2)}$, $h_{(1)}$, $f^a_{(2)}$ and
$f^h_{(1)}$ are not determined and remain as free parameters.
The higher-order coefficients ($n \ge 3$) are determined
in terms of these parameters.
Four free parameters are determined by boundary conditions at infinity.
The finiteness of action requires
$a$, $h$ $\rightarrow 0$ faster than $1/r^2$ at infinity. This
condition also requires $f_a$, $f_h$ $\rightarrow 0$.

We have introduced $\rho$ as a free scale parameter.
We adjust this parameter $\rho$ so that $a_{(2)}=-2$
to make the radius of the valley instanton unity.
As a result we have four parameters $h_{(1)}$, $f^a_{(2)}$,
$f^h_{(1)}$ and $\rho v$ for a given $\nu$.
These four parameters are determined so that $a$, $h$, $f^a$,
and $f^h$ $\rightarrow 0$ at infinity.

\begin{figure}
\centerline{
\epsfxsize=13cm
\epsfbox{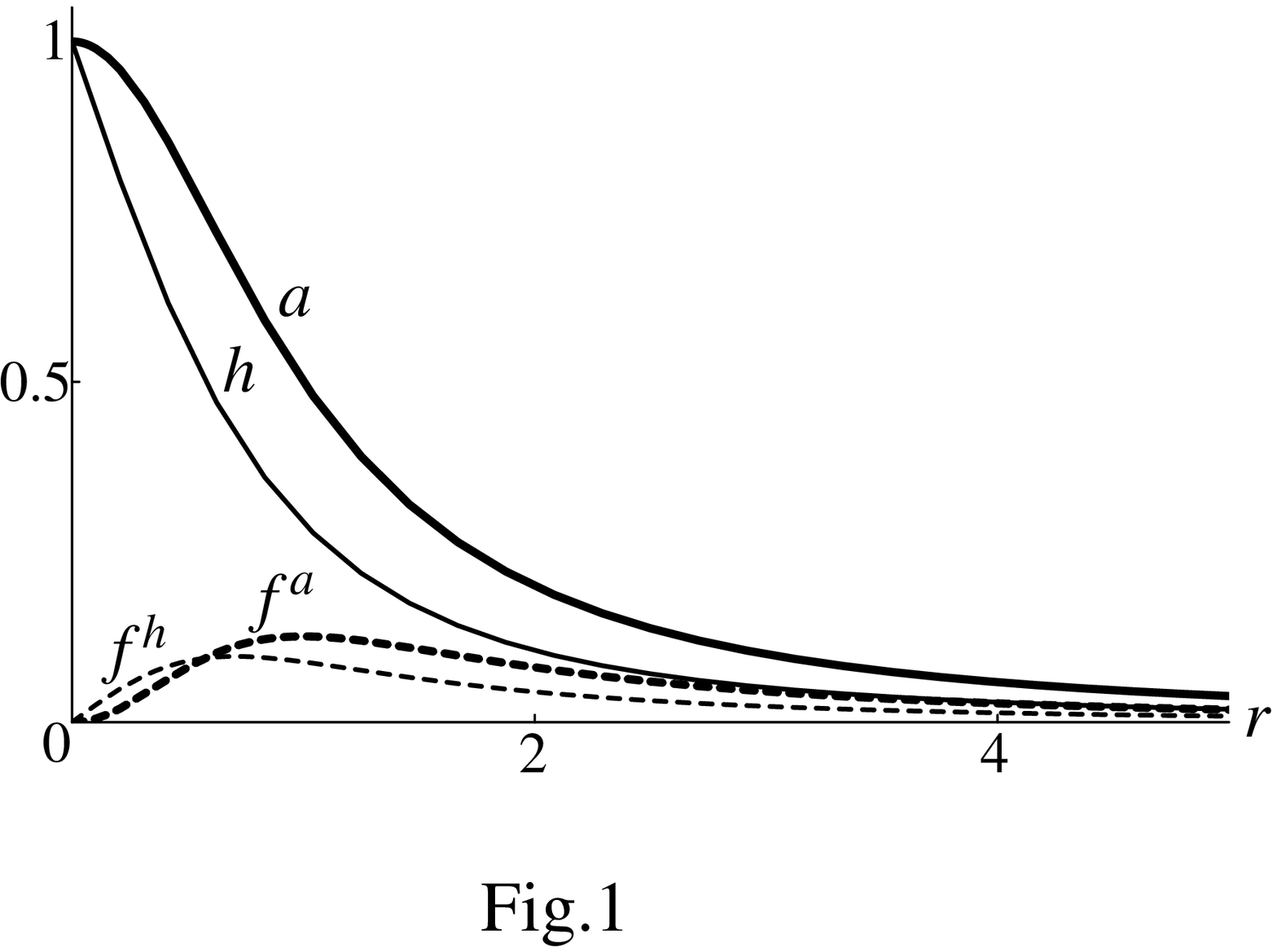}
}
\caption{Shapes of the numerical solution of $a(r)$, $f^{a}(r)$,
$h(r)$ and $f^{h}(r)$
for $\rho v =0.1$ near the origin.}
\label{config1}
\end{figure}

\begin{figure}
\centerline{
\epsfxsize=13cm
\epsfbox{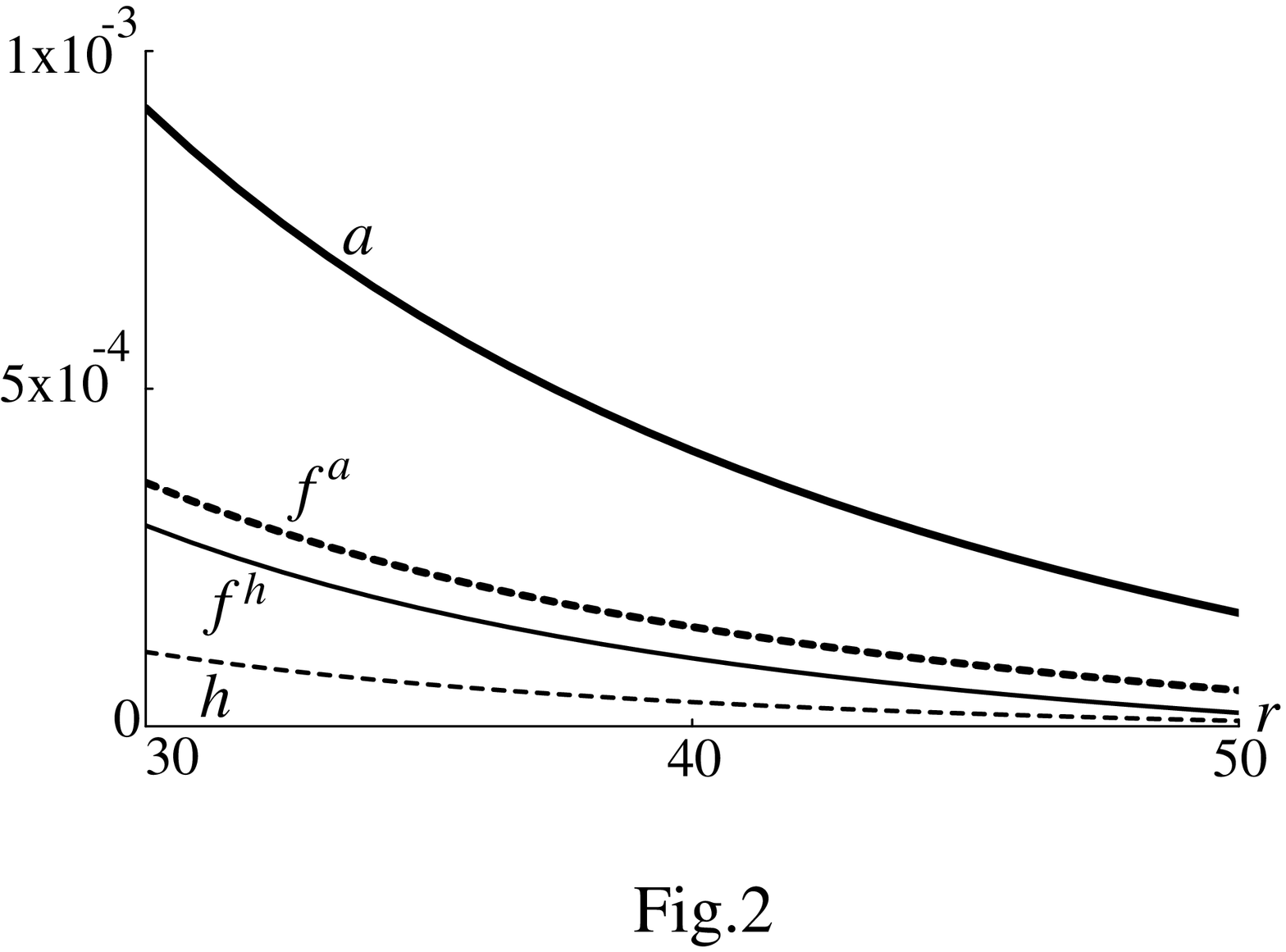}
}
\caption{Shapes of the numerical solution of $a(r)$, $f^{a}(r)$,
$h(r)$ and $f^{h}(r)$
for $\rho v =0.1$ in the asymptotic region.}
\label{config2}
\end{figure}

A numerical solution is plotted in Fig.\ref{config1} and
Fig.\ref{config2} for $\rho=0.1$ and $\lambda /g^2 =1$.
Fig.\ref{config1} is for the region near the origin and
 Fig.\ref{config2} is for the asymptotic region.
This behavior of the numerical solution agrees with the result of the
previous section.
In fact, if we plot the numerical solution and
the instanton solution (\ref{eq:a0}) with $C=1/4$ near the origin,
these two lines completely overlap with each other at the present scale.
In the same way, the solution in Fig.\ref{config2} and
the analytic solution (\ref{eqn:aa})-(\ref{eqn:afh}) also overlap
with each other.

\begin{figure}
\centerline{
\epsfxsize=13cm
\epsfbox{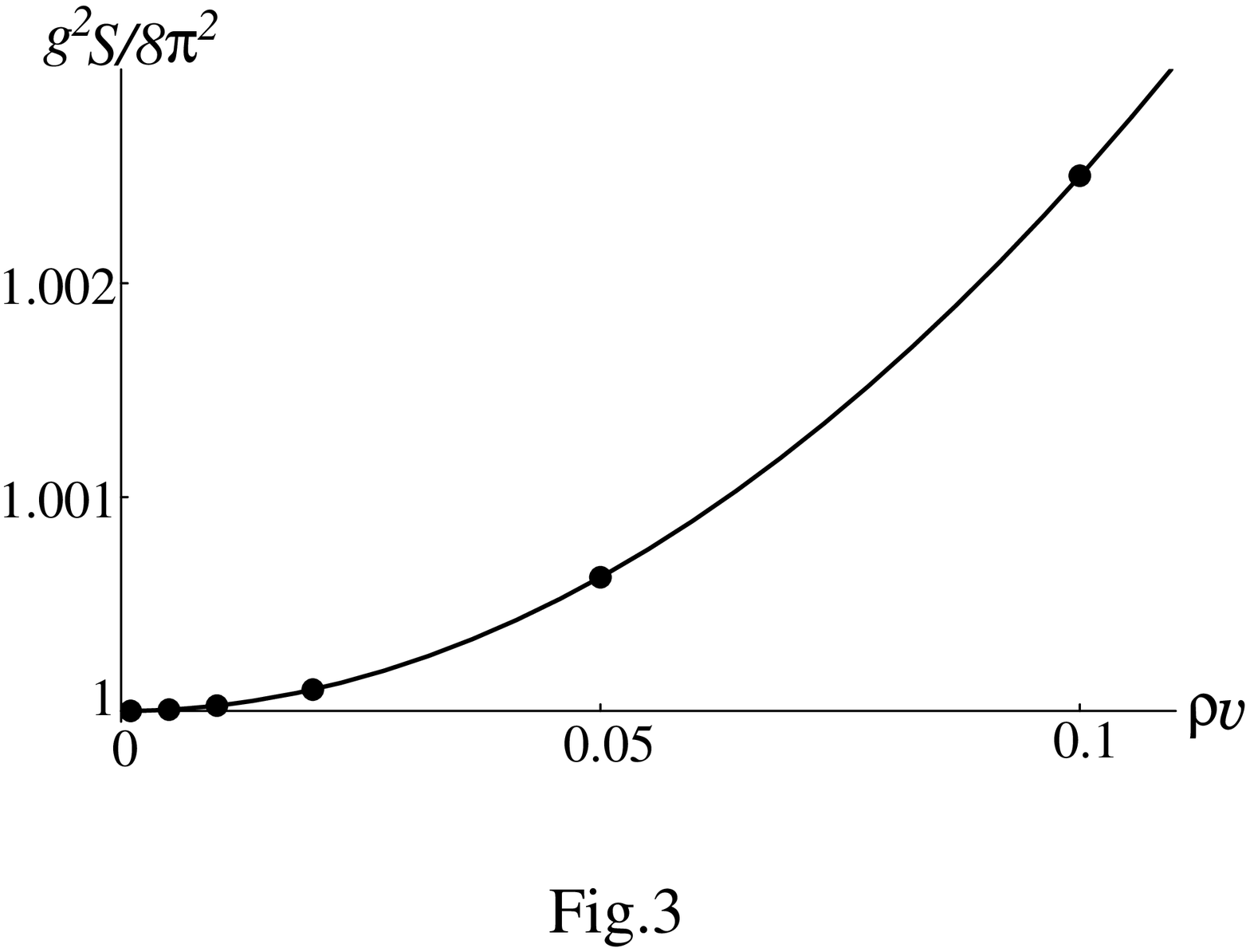}
}
\caption{The action $S$ (in units $g^2 S/8 \pi^2$) of the numerical solution of
the valley equation, at $\lambda /g^2 =1$, as a function of the
parameter $\rho v$. The solid line is the behavior of the analytical result
that $g^2 S/8 \pi^2 = 1 + \rho^2/4$.}
\label{fig:action}
\end{figure}

The values of the action at $\rho v=0.001$, $0.005$, $0.01$, $0.02$,
$0.05$ and $0.1$ are plotted in Fig.\ref{fig:action}.
The solid line shows the behavior of the action of the analytical result
(\ref{eqn:aaction}).
This figure shows that our numerical solutions are quite consistent
 with (\ref{eqn:aaction}).

\begin{figure}
\centerline{
\epsfxsize=13cm
\epsfbox{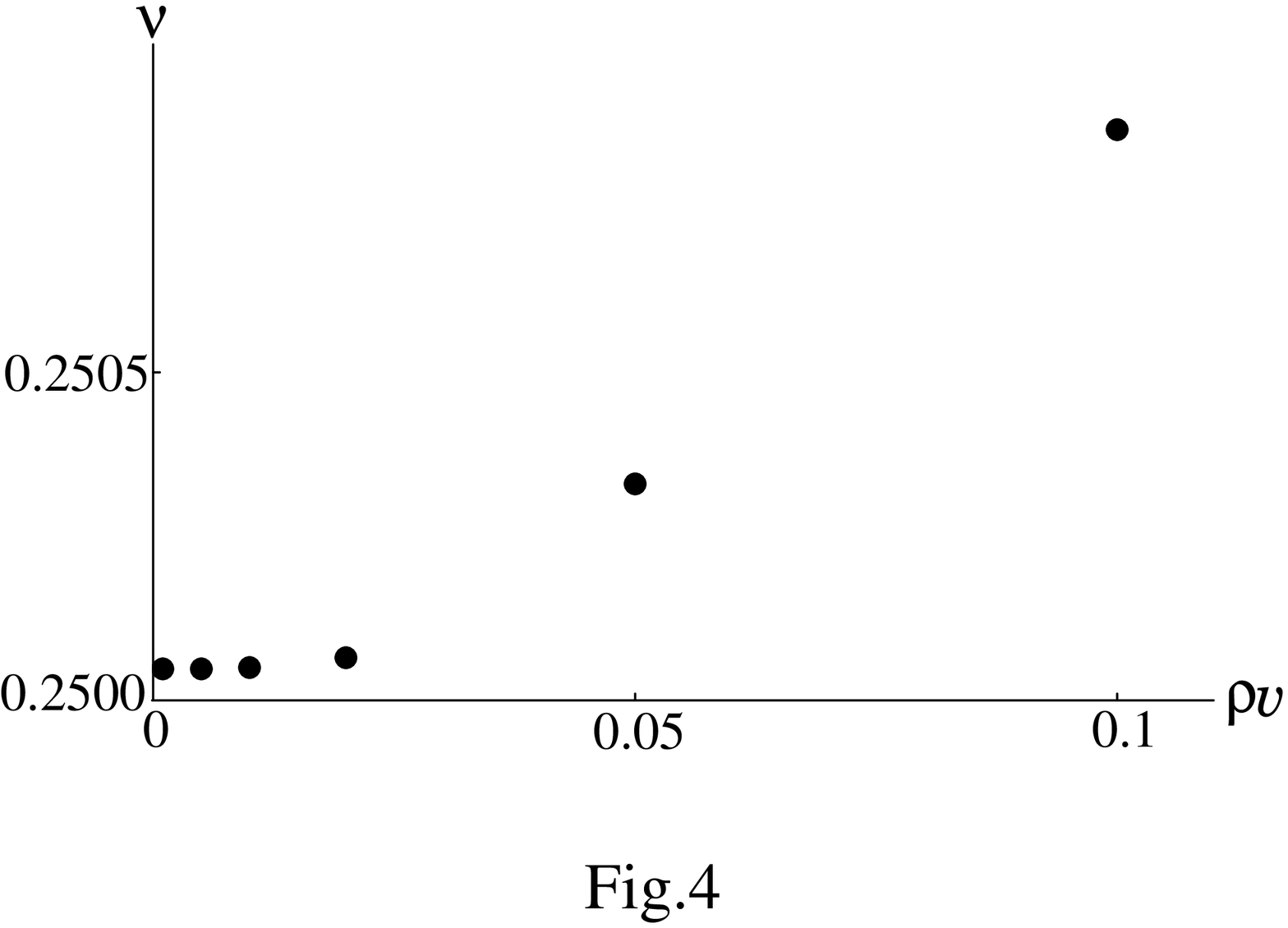}
}
\caption{Numerical results of the scaled eigenvalues $\nu$ at
$\rho v=0.001$, $0.005$, $0.01$, $0.02$, $0.05$ and $0.1$.}
\label{fig:eigen}
\end{figure}

The relation between the scaled eigenvalue $\nu$ and $\rho v$ is plotted in
Fig.\ref{fig:eigen}.
We find the tendency that $\nu \to 1/4$ as $\rho v \to 0$.
(For very small $\rho v$, the values of $\nu$ are different from 1/4.
This, however, should not be taken seriously due to the numerical
difficulties.)
This result also confirms the analysis of section 3.

We summarize all the numerical data in the Table \ref{tab:allresult}.

\section{Conclusions}
In this letter we have presented a construction of the
finite size instanton in the SU(2) gauge-Higgs system.
We used the new valley method for this purpose.
{}From the analysis both analytic and numerical, we have
established the existence of the solution, which has
the same tensor structure as the point-like instanton,
but has a finite radius.

We have found that the deformation of the valley instanton is
smaller than that of the constraint instanton.
For example, there appears a $O((\rho v)^{2})$
correction to $a(x)$ in the constraint instanton \cite{Aff}, while
there is no $O((\rho v)^{2})$ correction in the valley instanton.
This infers that the behavior of the large-radius valley instanton
is quite different from that of the constrained instanton
We expect that the valley instanton
has a much smaller action than the constrained instanton,
although analysis of such a configuration is
difficult both analytically and numerically.
Instead, we have carried out analysis of the scalar
field case studied in \cite{Aff} and found that
in fact the valley instanton has much more desirable behaviors.
Thus we expect that overall our valley instanton
has larger contribution to the functional integral
and gives us a better approximation scheme.

We note here that there is a way to incorporate
the new valley equation itself in the constrained instanton
formalism: As noted in \cite{aw}, the new valley equation
can be understood as the equation of motion under the constraint
$\int d^4x \sum_\alpha (\delta S / \delta \phi_\alpha (x))^2$.
(The eigenvalue $\lambda$ plays the role of the constraint
parameter.)
In this sense, the new valley method gives us
the definition of the constraint:
There is no arbitrariness in the constraint as we mentioned
in the introduction. The constraint determined
by the new valley method guarantees the effectiveness
of the approximation scheme, for the new valley method
has many virtues as a means of determining the dominant
configurations.

In this letter, we have not incorporated fermions in the theory.
Its introduction is straightforward, however.
We need to solve the mode equations for fermion
fields in the background of the valley instanton, as
we have no reason to take into account the
fermion contribution to the new valley equation itself.
Analysis along this line is in progress and will be
reported in future publications.

\vskip1cm
\centerline{\large\bf Acknowledgment}

\noindent
One of the authors (H.A.) is supported in part by the
Grant-in-Aid \#C-07640391 from the Ministry of Education, Science and Culture.
M.~S. and S.~W. are
the fellows of the Japan Society for the Promotion of
Science for Japanese Junior Scientist.

\newpage

\newcommand{\J}[4]{{\sl #1} {\bf #2} (19#3) #4}
\newcommand{\MPL}{Mod.~Phys.~Lett.}
\newcommand{\NP}{Nucl.~Phys.}
\newcommand{\PL}{Phys.~Lett.}
\newcommand{\PR}{Phys.~Rev.}
\newcommand{\PRL}{Phys.~Rev.~Lett.}
\newcommand{\AP}{Ann.~Phys.}
\newcommand{\CMP}{Commun.~Math.~Phys.}
\newcommand{\CQG}{Class.~Quant.~Grav.}
\newcommand{\PRP}{Phys.~Rept.}
\newcommand{\SPU}{Sov.~Phys.~Usp.}
\newcommand{\RMPA}{Rev.~Math.~Pur.~et~Appl.}
\newcommand{\SPJ}{Sov.~Phys.~JETP}
\newcommand{\MP}{Int.~Mod.~Phys.}

\newpage
\newcommand{\namelistlabel}[1]{\mbox{#1}\hfil}
\newenvironment{namelist}[1]{%
 \begin{list}{}
        {\let\makelabel\namelistlabel
        \settowidth{\labelwidth}{#1}
        \setlength{\leftmargin}{1.1\labelwidth}}
}{%
\end{list}}

\begin{Large}
\begin{flushleft}
{\bf Figure captions}
\end{flushleft}
\end{Large}
\begin{namelist}{Fig. 1}
\item[Fig.1]Shapes of the numerical solution of $a(r)$, $f^{a}(r)$,
$h(r)$ and $f^{h}(r)$
for $\rho v =0.1$ near the origin.
\item[Fig.2]Shapes of the numerical solution of $a(r)$, $f^{a}(r)$,
$h(r)$ and $f^{h}(r)$
for $\rho v =0.1$ in the asymptotic region.
\item[Fig.3]The action $S$ (in units $g^2 S/8 \pi^2$) of the numerical solution
of the valley equation, at $\lambda /g^2 =1$, as a function of the
parameter $\rho v$. The solid line is the behavior of the analytical result
that $g^2 S/8 \pi^2 = 1 + \rho^2/4$.
\item[Fig.4]Numerical results of the scaled eigenvalues $\nu$ at
$\rho v=0.001$, $0.005$, $0.01$, $0.02$, $0.05$ and $0.1$.
\end{namelist}

\newpage
\begin{table}[p]
\begin{center}
\doublerulesep=0pt
\def\arraystretch{1.3}
\begin{tabular}{|c|c|c|c|c|c|}\hline
{$\rho v$ } & {$\nu$} & $h_{(1)}$ & $f^a_{(2)}$ &
$f^h_{(1)}$ & $g^{2}S/8\pi^2$ \\
\hline
0.001 & 0.250048 & -1.000065 & 0.500064 & 0.250048 & 1.000000 \\
\hline
0.005 & 0.250048 & -1.000069 & 0.500069 & 0.250046 & 1.000006 \\
\hline
0.01 & 0.250050 & -1.000010 & 0.500108 & 0.250041 & 1.000025 \\
\hline
0.02 & 0.250065 & -1.000130 & 0.500289 & 0.250130 & 1.000100 \\
\hline
0.05 & 0.250330 & -1.000200 & 0.501419 & 0.250519 & 1.000625 \\
\hline
0.1 & 0.250870 & -1.000489 & 0.504364 & 0.251544 & 1.002503 \\
\hline
\end{tabular}
\end{center}
\caption{The numerical data of $\rho v$, $\nu$, $h_{(1)}$, $f^a_{(2)}$,
$f^h_{(1)}$ and $g^{2}S/8\pi^2$.}
\label{tab:allresult}
\end{table}

\newpage

\begin{thebibliography}{99}
\bibitem{thooft} G.~'t Hooft, \J{\PRL}{37}{76}{8};
\J{\PR}{D14}{76}{3422}.
\bibitem{manton} F.~R.~Klinkhamer and N.~S.~Manton, \J{\PR}{D30}{84}{2212}.
\bibitem{Ring} A.~Ringwald, \J{\NP}{B330}{90}{1}.
\bibitem{Esp} O.~Espinosa \J{\NP}{B343}{90}{310}.
\bibitem{AK} H.~Aoyama and H.~Kikuchi, \J{\PL}{B247}{90}{75};
\J{\PR}{D43}{91}{1999}; \J{\MP}{A7}{92}{2741}.
\bibitem{NSVZ} V.~A.~Novikov, M.~A.~Shifman, A.~I.~Vainshtein and
V.~I.~Zakharov, \J{\NP}{B229}{83}{407}.
\bibitem{ADS} I.~Affleck, M.~Dine and N.~Seiberg, \J{\NP}{B241}{84}{493}.
\bibitem{Yung} A.~V.~Yung, \J{\NP}{B297}{88}{47}.
\bibitem{AKMRV} D.~Amati, K.~Konishi, Y.~Meurice, G.~C.~Rossi and
G.~Veneziano, \J{\PRP}{162}{88}{169} and references therein.
\bibitem{Aff} I.~Affleck, \J{\NP}{B191}{81}{429}.
\bibitem{KR} V.~V.~Khoze and A.~Ringwald, \J{\NP}{B355}{91}{351}.
\bibitem{AKnv} H.~Aoyama and H.~Kikuchi, \J{\NP}{B369}{92}{219}.
\bibitem{aw} H.~Aoyama and S.~Wada, \J{\PL}{B349}{95}{279}
\bibitem{balyun} I.~I.~Balitsky and A.~V.~Yung, \J{\PL}{B168}{86}{113}
\end{thebibliography}
\end{document}